\documentclass[pre,twocolumn,amsmath,amssymb,nofootinbib,floatfix,superscriptaddress]{revtex4}

\usepackage{graphicx, bm, xcolor, bbold, enumerate, float}
\usepackage[normalem]{ulem}
\usepackage[breaklinks]{hyperref}

\makeatletter
\def\graphicscale{\twocolumn@sw{0.3}{0.4}}
\def\graphicthreescale{\twocolumn@sw{0.3}{0.4}}

\begin{document}

\title{Out-of-equilibrium spinodal-like scaling behaviors at the
  thermal \\ first-order transitions of three-dimensional $q$-state
  Potts models}
  
\author{Andrea Pelissetto}
\altaffiliation{Authors are listed in alphabetic order.}
\affiliation{Dipartimento di Fisica dell'Universit\`a di Roma
  La Sapienza and INFN, Sezione di Roma I, I-00185 Roma, Italy}

\author{Davide Rossini}
\altaffiliation{Authors are listed in alphabetic order.} 
\affiliation{Dipartimento di Fisica dell'Universit\`a di Pisa
  and INFN, Largo Pontecorvo 3, I-56127 Pisa, Italy}

\author{Ettore Vicari}
\altaffiliation{Authors are listed in alphabetic order.}
\affiliation{Dipartimento di Fisica dell'Universit\`a di Pisa,
  Largo Pontecorvo 3, I-56127 Pisa, Italy}

\date{\today}

\begin{abstract}
  We study the out-of-equilibrium spinodal-like dynamics of
  three-dimensional $q$-state Potts systems driven across their
  thermal first-order transition in the thermodynamic limit, by a
  relaxational (heat-bath) dynamics. During the evolution, the inverse
  temperature $\beta$ increases linearly with time, as $\delta\beta(t)
  \equiv \beta(t)- \beta_{\rm fo} \sim t/t_s$, where $\beta_{\rm fo}$
  is the inverse temperature at the transition point, $t$ is the time
  and $t_s$ is a time scale.  The dynamics starts at $t_i< 0$ from an
  ensemble of disordered configurations equilibrated at inverse
  temperature $\beta(t_i)<\beta_{\rm fo}$ and ends at positive values
  of $t$, when the system is ordered (this is analogous to a standard
  Kibble-Zurek protocol).  The time-dependent energy density shows an
  out-of-equilibrium scaling behavior in the large-$t_s$ limit, in
  terms of the scaling variable $t(\ln t)^\kappa/t_s$. The
  corresponding exponent turns out to be consistent with $\kappa=3/2$
  (with a good accuracy), which is the value obtained by assuming that
  the initial nucleation of ordered regions provides the relevant
  mechanism for the passage from one phase to the other.  The scaling
  behavior implies a spinodal-like phenomenon close to the transition
  point: the passage from the disordered to the ordered phase,
  composed of large ordered regions of different color, occurs at
  $\delta\beta(t)=\delta\beta_*>0$, where $\delta\beta_*$ decreases as
  $1/(\ln t_s)^{3/2}$ in the large-$t_s$ limit.
\end{abstract}

\maketitle

\section{Introduction}
\label{intro}

Statistical systems driven across phase transitions exhibit
out-of-equilibrium behaviors, even when the system parameters are
varied very slowly, because large-scale modes are not able to
equilibrate on any time scale in the thermodynamic limit. Notable
examples of out-of-equilibrium phenomena at continuous and
discontinuous transitions are hysteresis, spinodal and coarsening
phenomena, critical aging, out-of-equilibrium finite-size scaling,
Kibble-Zurek (KZ) defect production. Such effects are observed in a
wide range of classical and quantum contexts (see, e.g.,
Refs.~\cite{Kibble-80,Binder-87,Bray-94,Zurek-96,CA-99, CG-04,PSSV-11,
  Biroli-16,RV-21,PV-24} for reviews, and also
Refs.~\cite{Zurek-85,CDTY-91,BCSS-94,RTMS-94,BBFGP-96,
  Ruutu-etal-96,MM-00,CPK-00,CGMB-01,MMR-02,
  MPK-03,ZDZ-05,CGM-06,MMARK-06,SHLVS-06,PG-08,WNSBDA-08,LFGC-09,
  GPK-10,NIW-11,CWCD-11,CEGS-12,
  Chae-etal-12,MBMG-13,EH-13,Ulm-etal-13,Pyka-etal-13,LDSDF-13,
  ICA-14, Corman-etal-14,PV-15, NGSH-15,Braun-etal-15,PV-16,DWGGP-16,
  PV-17,PV-17b,LZ-17,PPV-18,SW-18,PRV-18,PRV-18-def,LZW-19,RV-20,
  PRV-20,DRV-20,CCP-21,SCD-21,CCEMP-22,TV-22,TS-23,DV-23,Surace-etal-24,
  PRV-25,PV-25,PRV-25b}).

In this paper, we study the out-of-equilibrium
spinodal-like\footnote{In the standard theory of first-order
transitions~\cite{Binder-87} one defines the spinodal line as the line
where the metastable state becomes unstable (the free-energy minimum
corresponding to the metastable state disappears). In the mean-field
approach, the spinodal line is located at an inverse temperature
$\beta_{\rm sp}$ that differs from the transition value $\beta_{\rm
  fo}$.  However, in short-range models there are no metastable states
for any $\delta\beta=\beta-\beta_{\rm fo}$ in the infinite-size
limit~\cite{Binder-87}.  From a dynamic perspective, in the limit of
very slow variations of the parameters in large-size
systems, it is not possibile to observe quasi-equilibrium metastable
states in short-range model when $\delta\beta$ is finite.
Spinodal-like behaviors can instead be observed in an appropriate
out-of-equilibrium scaling regime~\cite{PV-17,PV-24,PRV-25b,PV-25}, as
also discussed in this paper. \label{footnote1}}
behavior observed when the temperature is slowly varied across a
thermal (i.e., driven by thermal fluctuations) first-order transition
(FOT) in the thermodynamic limit. We focus on the three-dimensional
(3D) $q$-state Potts model evolving under a purely relaxational
dynamics and study the out-of-equilibrium scaling behavior developed
when the system is driven across the FOT at $\beta = \beta_{\rm fo}$
in the thermodynamic limit.  For this purpose, we consider a standard
KZ protocol in which the inverse temperature varies linearly across
the FOT as $\delta\beta(t) = \beta(t)-\beta_{\rm fo}\sim t/t_s$, where
$t_s$ is the time scale of the variation of $\beta$.  The evolutions
start from an ensemble of configurations equilibrated at
$\beta<\beta_{\rm fo}$ in the disordered phase and ends at a positive
value ot $t$, when the system is in the ordered phase.

Out-of-equilibrium scaling behaviors arising from analogous KZ
protocols have been extensively investigated at continuous transitions
(see, e.g., Refs.~\cite{Kibble-80,Zurek-85,RV-21,Zurek-96,ZDZ-05,PG-08,
  PSSV-11,CEGS-12,TV-22,DV-23}). In this context, such protocols lead
to characteristic out-of-equilibrium scaling laws and, in particular,
to the distinctive formation of defects~\cite{Kibble-80,Zurek-96},
which are controlled by the universal critical exponents
of the underlying equilibrium transition.  KZ, or more general quenching
protocols, have also been studied at first-order classical and quantum
transitions~\cite{Binder-87,Pfleiderer-05,RV-21,PV-24}, revealing more
complex out-of-equilibrium behaviors (see, e.g.,
Refs.~\cite{RV-21,PV-24,MM-00,LFGC-09,NIW-11,
  ICA-14,PV-15,PV-16,PV-17,PV-17b,LZ-17,PPV-18,SW-18,PRV-18,
  PRV-18-def,LZW-19,PRV-20,DRV-20,CCP-21,SCD-21,
  CCEMP-22,TS-23,Surace-etal-24,PRV-25}).  In particular, qualitatively
different mechanisms work at FOTs in a finite volume and in the
thermodynamic limit, potentially leading to unrelated scaling
behaviors~\cite{PV-25,PRV-25b}. While the out-of-equilibrium
finite-size scaling theory of the dynamics across FOTs is well
established and supported by numerical
studies~\cite{PV-16,PV-17,PPV-18,PRV-18-def,RV-21,PV-24,PRV-25}, the
scaling behavior in the thermodynamic limit exhibits distinct and
less understood peculiar features.

Our study is meant to extend our understanding of the phenomenology of
out-of-equilibrium spinodal-like phenomena occurring in short-ranged
statistical models. This issue has already been explored in other
contexts, including the KZ dynamics at magnetic low-temperature FOTs
in Ising systems~\cite{PV-25,PRV-25b}, and at thermal FOTs in
two-dimensional (2D) Potts models~\cite{PV-17}.  These studies reveal
the emergence of spinodal-like scaling behaviors in the thermodynamic
limit. Indeed, the observables characterizing the FOT (such as the
energy density at thermal FOTs or the magnetization at magnetic FOTs)
exhibit time-dependent behaviors that scale in terms of the
variable~\cite{PV-17,PV-24,PRV-25b,PV-25}
\begin{equation}
  \sigma(t,t_s) = { t \,(\ln t)^\kappa \over t_s}.
  \label{sdef}
\end{equation}
The exponent $\kappa$ turns out to depend on the spatial dimension $d$
of the model and on the nature of the first-order transition (i.e.,
whether it is driven by thermal or magnetic perturbations). The
determination of the value of $\kappa$ requires identifying the
dynamic mechanism which provides the longest time scale characterizing
the passage across the transition.  One possibility is that the
slowest mode is associated with the nucleation of smooth droplets,
leading to the prediction $\kappa=d/(d-1)$, as discussed below.
Behaviors consistent with this prediction have been observed at the
thermal FOTs of 2D Potts models and at low-temperature magnetic FOTs
of 2D Ising models.  However, at magnetic FOTs in higher-dimensional
Ising systems, the observed scaling shows values of $\kappa$ 
different  from the droplet prediction $\kappa=d/(d-1)$: one finds
$\kappa=1$ and $\kappa=1/2$ for 3D and 4D Ising systems, respectively.
There results suggest the presence of a different, not yet understood,
mechanism governing the long-time behavior when the system is driven
across magnetic FOTs.  We believe that extending the known
phenomenology of the behaviors at FOTs is essential for achieving a
deeper understanding of the out-of-equilibrium mechanisms
characterizing phase changes at first-order classical and quantum
transitions.

Here we consider the 3D $q$-state Potts model and investigate its
dynamic behavior as the system is driven across its thermal FOT, with
the purpose of identifying the physical mechanism that controls the
slowest time scale of the dynamics.  In particular, we wish to
understand whether the formation of stable smooth droplets is indeed
the relevant mechanism, as appears to be the case in all 2D systems
studied so far, or whether a slower, and currently not fully
understood mechanism is at work, as observed in 3D and 4D Ising
models.  We perform Monte Carlo (MC) simulations for the $q=6$ and
$q=10$ Potts models on large lattices, allowing us to determine the
dynamical behavior in the infinite-volume limit.  Our numerical MC
data exhibit scaling in terms of the variable~\eqref{sdef} with
$\kappa = 3/2$. This value agrees with the droplet prediction,
confirming that the relevant mechanism for the spinodal-like scaling
behavior is the nucleation of smooth droplets.

The paper is organized as follows.  In Sec.~\ref{moddyn} we define 
the 3D $q$-state Potts model and the KZ protocol.
Moreover, we outline the argument based on droplet nucleation that
allows us to predict the characterizing features of the
out-of-equilibrium scaling behavior in the thermodynamic limit. In
Sec.~\ref{numres} we report our numerical analyses for 
$q=6$ and $q=10$.  Finally, in Sec.~\ref{conclu} we draw our conclusions.

\section{3D Potts models and dynamic protocol}
\label{moddyn}

\subsection{The model}
\label{model}

We consider the 3D nearest-neighbor $q$-state Potts
model~\cite{Wu-82}. The Hamiltonian and partition function are
\begin{equation}
  H = J \sum_{\langle {\bm x}{\bm y}\rangle} \left[1-\delta(s_{{\bm
        x}}, s_{ {\bm y}})\right],\qquad
  Z = \sum_{\{s_{\bm x}\}} e^{-\beta H}, 
  \label{potts}
\end{equation}
where $\beta=1/T$, the sum in $H$ is over the nearest-neighbor sites
of a cubic lattice, $s_{\bm x}$ are integer variables $s_{{\bm x}}=1,
2, \ldots, q$, $\delta(a,b)=1$ if $a=b$ and zero otherwise.  We
consider systems of size $L$ in each direction with periodic boundary
conditions, which preserve the $q$-state permutation symmetry.
Moreover, we set $J=1$ without loss of generality.

The 3D Potts model undergoes a FOT at a finite $\beta=\beta_{\rm fo}$
for any $q\ge 3$, between a disordered phase and $q$ equivalent
ordered phases, driven by thermal fluctuations~\cite{Wu-82,BBD-08}.
The energy density plays the same role as the magnetization at
magnetic transitions, while $\delta \beta = \beta - \beta_{\rm fo}$
acts as the corresponding magnetic field. We consider the rescaled
energy density,
\begin{equation}
  E(\beta)\equiv {q\over 3(q-1)} {\langle H \rangle\over V} ,
  \qquad V = L^3,
  \label{ener}
\end{equation}
which decreases monotonically from $E=1$ (for $\beta\to 0$)
to $E=0$ (for $\beta\to\infty$).  The rescaled energy density $E(\beta)$
is discontinuous at $\beta_{\rm fo}$ in the thermodynamic limit, i.e.,
\begin{equation}
  {\rm lim}_{\beta\to\beta_{\rm fo}^{\pm}}  \; {\rm lim}_{V\to\infty} \;
  E(\beta) = E_{\pm}, 
  \label{deltae}
  \end{equation}
leading to the rescaled latent heat 
\begin{equation}
  \Delta_h \equiv  E_+ - E_- > 0.
  \label{deltah}
\end{equation}
Therefore, energy-density values $E>E_+$ indicate that the system is
in the disordered phase, and $E<E_-$ in the low-temperature ordered
phase.

We finally remark that, in the absence of perturbations that break the
$q$-state permutation symmetry, which is also preserved by 
periodic boundary conditions, the magnetization
\begin{equation}
  M_k = {1\over L^2} \langle \sum_{\bm x} \mu_k({\bm x}) \rangle,
  \qquad \mu_k({\bm x}) \equiv {q \delta(s_{\bm x},k) - 1\over q-1},
  \label{mkdef}
\end{equation}
vanishes for any $\beta$.  Therefore, for periodic boundary
conditions and in the absence of symmetry-breaking perturbations,
sufficiently large systems in the low-temperature phase 
exhibit a coexistence of large ordered regions associated with different $q$
states, to ensure a vanishing magnetization.

\subsection{Kibble-Zurek protocols across first-order transitions}
\label{dynprot}

To study the out-of-equilibrium spinodal-like behavior induced by
driving the system across the thermal FOTs, we consider a KZ
protocol in which the inverse temperature $\beta$ is varied linearly
across the transition at $\beta_{\rm fo}$, as 
\begin{equation}
  \delta(t)\equiv {\delta\beta(t)\over \beta_{\rm
      fo}}={\beta(t)\over \beta_{\rm fo}} - 1 = {t\over t_s},
  \label{deltat}
\end{equation}
where $t_s$ is a time scale.  The dynamics starts at a time $t_i<0$,
from an ensemble of equilibrium configurations at the inverse
temperature $\beta_i=\beta_{\rm fo} \left(1 + t_i/t_s\right) <
\beta_{\rm fo}$ in the high-temperature phase, and ends at $t_f > 0$,
thus $\beta_f>\beta_{\rm fo}$, in the low-temperature phase. This
protocol is general, as any smooth time dependence can be approximated
by a linear function around $\beta_{\rm fo}$.  We consider a heat-bath
dynamics~\cite{Binder-76,Creutz-book} (at each site, a new spin is
chosen using the conditional probability distribution at fixed
neighboring spins), which is a specific example of a purely
relaxational dynamics (model-A dynamics in the standard
terminology~\cite{HH-77}).  We study the KZ dynamics in the
thermodynamic limit. First, we determine the infinite-size limit
keeping $t_s$ fixed, then we study the scaling behavior in the
large-$t_s$ limit.  Information on the dynamics is provided by the
average energy density as a function of time,
\begin{equation}
  E(t,t_s,L)= {q\over 3(q-1)} {\langle H \rangle_t\over V},
  \label{etts}
\end{equation}
where the average $\langle H \rangle_t$ at time $t$ is performed over
a large number of trajectories starting from thermalized
configurations at inverse temperature $\beta_i$. Therefore, the
average is performed over both the initial equilibrium ensemble at
$\beta_i$ and the stochastic trajectories arising from the heat-bath
relaxational dynamics. In practice, we consider one trajectory for
each initial equilibrium configuration, so all trajectories that we
analyze are independent if the initial configurations are
decorrelated.

\begin{figure}[tbp]
  \includegraphics[width=0.95\columnwidth, clip]{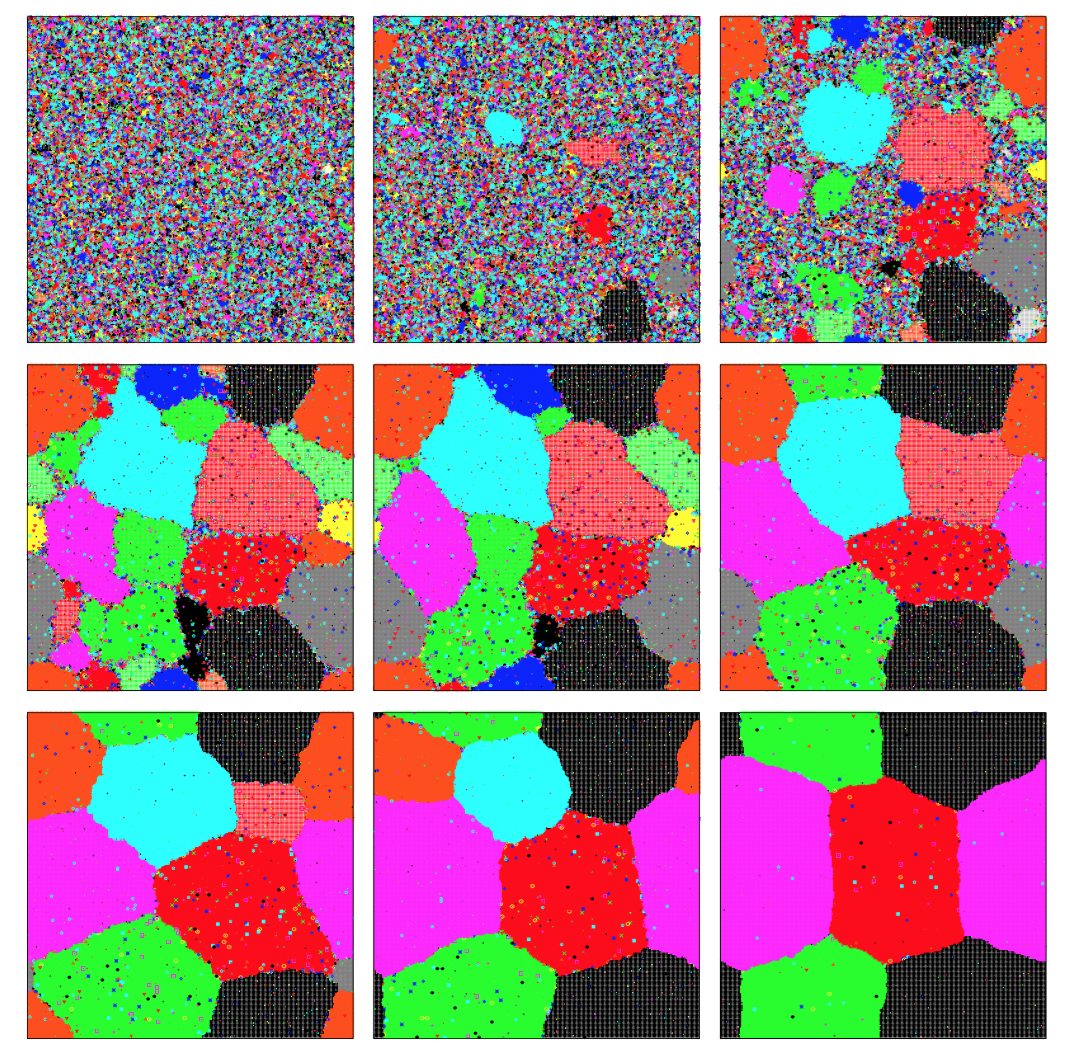}
  \caption{Snapshots of the configurations of a 2D $q=20$ Potts system
    of size $L=512$, driven across its FOT by a KZ relaxational
    dynamics (time, and therefore $\beta$, increases from left to
    right and from top to bottom).  The images show the nucleation of
    different ordered regions, which grow as time increases.
    Different colors correspond to the $q$ distinct states. These
    snapshots should give an idea of what happens in 3D Potts models
    as well.}
  \label{2dsnapshots}
\end{figure}

We finally note that the time-dependent magnetization defined in
Eq.~(\ref{mkdef}) always vanishes, because both the initial
equilibrium ensemble of states and the out-of-equilibrium relaxational
dynamics preserve the $q$-state permutation symmetry.  Therefore, to
maintain a vanishing magnetization, the final stage of the KZ dynamics
in the low-temperature phase is characterized by the growth of ordered
regions associated with the $q$ different Potts states (see, for
example, Fig.~\ref{2dsnapshots} showing snapshots obtained for the 2D
Potts model~\cite{PV-17,PV-24}).  The behavior in three dimensions is
qualitatively analogous, with the growth of 3D $q$-state ordered
regions.

\subsection{Spinodal-like scaling  behavior in the thermodynamic limit}
\label{TLsca}

Previous studies of the out-of-equilibrium KZ dynamics across
classical and quantum FOTs in the thermodynamic
limit~\cite{PV-17,PV-24,PV-25,PRV-25,PRV-25b} have shown that the
relevant observables develop an asymptotic spinodal-like scaling
behavior in the large-$t_s$ limit, in terms of the scaling
variable~\eqref{sdef}.  The exponent $\kappa$ apparently depends on
the spatial dimension $d$ of the model and on the nature of the FOT,
i.e., whether it is driven by thermal or magnetic perturbations.
The exponent $\kappa$ is determined by the time dependence of the
slowest modes that control the passage across the transition. At FOTs,
one relevant time scale characterizing the transitions between the two
phases is the one associated with the nucleation of stable
droplets~\cite{Binder-87,RTMS-94}. If this time scale is the largest
one, we can predict the exponent $\kappa$ in Eq.~\eqref{sdef}.

The transition to the ordered phase starts with the
nucleation of droplets of the $q$ ordered phases, of size $R_{\rm dr}
\ll L$ for a large system size $L$.  We assume droplets to have a
smooth boundary, so the volume of each droplet is of order $R_{\rm
  dr}^d$, while the area scales as $R_{\rm dr}^{d-1}$.~\footnote{The
radius of the droplets must be larger than the correlation length of
the small-range fluctuations, which determine the intrinsic interface
thickness, otherwise more complex fractal structures should be also
taken into account (see, e.g., Ref.~\cite{Binder-87} for a discussion
of this issue).\label{footnote2}} The time $t$ needed to create a
droplet of size $R_{\rm dr}$ should increase exponentially with the
area $R_{\rm dr}^{d-1}$ of the droplet \cite{Binder-87,RTMS-94,PV-17},
so $\ln t \sim R_{\rm dr}^{d-1}$, leading to the relation
\begin{equation}
  R_{\rm dr}(t) \sim (\ln t)^{1/(d-1)}.
  \label{rtdep}
\end{equation}
Since all spins in the droplet have the same color, the energy of the
droplet is proportional to its volume $R_{\rm dr}^d$.  This suggests
that the basic scaling variable to describe the KZ dynamics at a
$d$-dimensional thermal FOT is
\begin{equation}
  \sigma(t,t_s) = R_{\rm dr}(t)^d \, \delta(t)
  = {t \, R_{\rm dr}(t)^d\over t_s},
  \label{sigmar}
\end{equation}
which quantifies the ordering energy with respect to the thermal
energy.  Therefore, under the hypothesis that the nucleation mechanism
of smooth droplets is the one providing the largest time scale, we end
up with Eq.~(\ref{sdef}), where $\kappa$ is given by
\begin{equation}
  \kappa = {d\over d-1}.
  \label{kappadro}
\end{equation}
The above scaling arguments imply the out-equilibrium scaling behavior
\begin{equation}
  E(t,t_s)\approx {\cal E}(\sigma),
  \label{etsca}
\end{equation}
for the energy density. The assumption that the nucleation mechanism
is the one controlling the scaling behavior requires that, once stable
droplets are created, the phase change occurs on a shorter time scale.
More precisely, if the typical phase-change time $\tau_{\rm pc}$ taken
by the system to go from one phase to the other satisfies $\tau_{\rm
  pc}/\tau_{\rm dr}\to 0$ for $t_s\to \infty$ at fixed $\sigma$, where
\begin{equation}
  \tau_{\rm dr} \sim {t_s \over(\ln t_s)^{\kappa}},
  \label{tsudr}
\end{equation}  
  is the typical time
needed to create the droplet in the KZ dynamics, the phase change is
effectively instantaneous in the scaling variable
$\sigma$.~\footnote{We assume that the phase change
starts at time $t_1$, when stable droplets appear in the
system---therefore $t_1 \approx \tau_{\rm dr}$---and ends at time $t_2
= t_1 + \tau_{\rm pc}$. If $\Delta\sigma = \sigma(t_2,t_s) -
\sigma(t_1,t_s)$, then we have $\Delta\sigma \approx {d\sigma\over dt}
\tau_{\rm pc} \approx (\ln t_1)^\kappa \tau_{\rm pc}/t_s =
\sigma(t_1,t_s) \tau_{\rm pc}/t_1$.  Thus, we have $\Delta\sigma\to 0$
if $\tau_{\rm pc}/\tau_{\rm dr}\to 0$.}
Thus, the scaling function ${\cal E}(\sigma)$ becomes
discontinuous at some value $\sigma = \sigma_*$, giving rise to a
spinodal-like behavior.

The assumption that droplet nucleation controls the dynamics of the
phase change turns out to be correct in two dimensions, both for the
thermal FOT of the Potts model~\cite{PV-17} and for the magnetic FOTs
of the Ising model. In both cases, numerical results exhibit excellent
scaling in terms of the variable~\eqref{sdef} with $\kappa = 2$, as
predicted by Eq.~\eqref{kappadro}.  In contrast, numerical
results~\cite{PV-25} for the magnetic FOTs of 3D and 4D Ising models
scale only if one adopts values of $\kappa$ that differ from
prediction~\eqref{kappadro}: one finds $\kappa\approx 1$ in three
dimensions and $\kappa\approx 1/2$ in four dimensions, clearly
different from the droplet predictions of Eq.~\eqref{kappadro},
$\kappa = 3/2$ and $\kappa = 4/3$ for $d=3$ and $d=4$, respectively.
These results suggest that a different mechanism determines the
largest time scale governing the passage from one phase to the other
in 3D and 4D Ising systems.  As we shall see, unlike 3D and 4D Ising
systems, the numerical analyses of the KZ dynamics at the thermal FOT
in the 3D Potts model yield $\kappa\approx 3/2$, consistently with the
prediction based on droplet nucleation.

We finally note that out-of-equilibrium scaling behaviors also appear
in the finite-size scaling limit, as discussed in Ref.~\cite{PV-17},
and numerically confirmed at the FOT of 2D Potts models. However, the
dynamical mechanisms relevant for finite-size systems differ from
those that are relevant in the thermodynamic limit, thus leading to
unrelated scaling behaviors~\cite{PRV-25b,PV-25}.  In this work, we do
not repeat the same finite-size scaling analysis for the 3D Potts
model, focusing instead only on the behavior in the thermodynamic
limit.

\section{Numerical results}
\label{numres}

We present numerical results obtained by MC simulations of the $q=6$
and $q=10$ 3D Potts models, for which~\cite{BBD-08} $\beta_{\rm fo} =
0.739214(3)$ and $\beta_{\rm fo}=0.881474(4)$, respectively. The
corresponding rescaled latent heats, defined in Eq.~\eqref{deltah},
are $\Delta_h\approx 0.4729$ for $q=6$ and $\Delta_h=0.6569$ for
$q=10$ (they are obtained using the results of Ref.~\cite{BBD-08}).
We analyze the out-of-equilibrium KZ behavior of the heat-bath
dynamics in the thermodynamic limit, defined as the limit $L\to
\infty$ while keeping $t_s$ fixed.  To this end, we performed MC
simulations at fixed $t_s$ for several system sizes, increasing $L$
until the average energy curves become approximately independent of
$L$.  We only show results for the energy density $E(t)$, as it
provides the most relevant information for the passage from the
high-temperature to the low-temperature phase.

\begin{figure}[tbp]
  \includegraphics[width=0.95\columnwidth, clip]{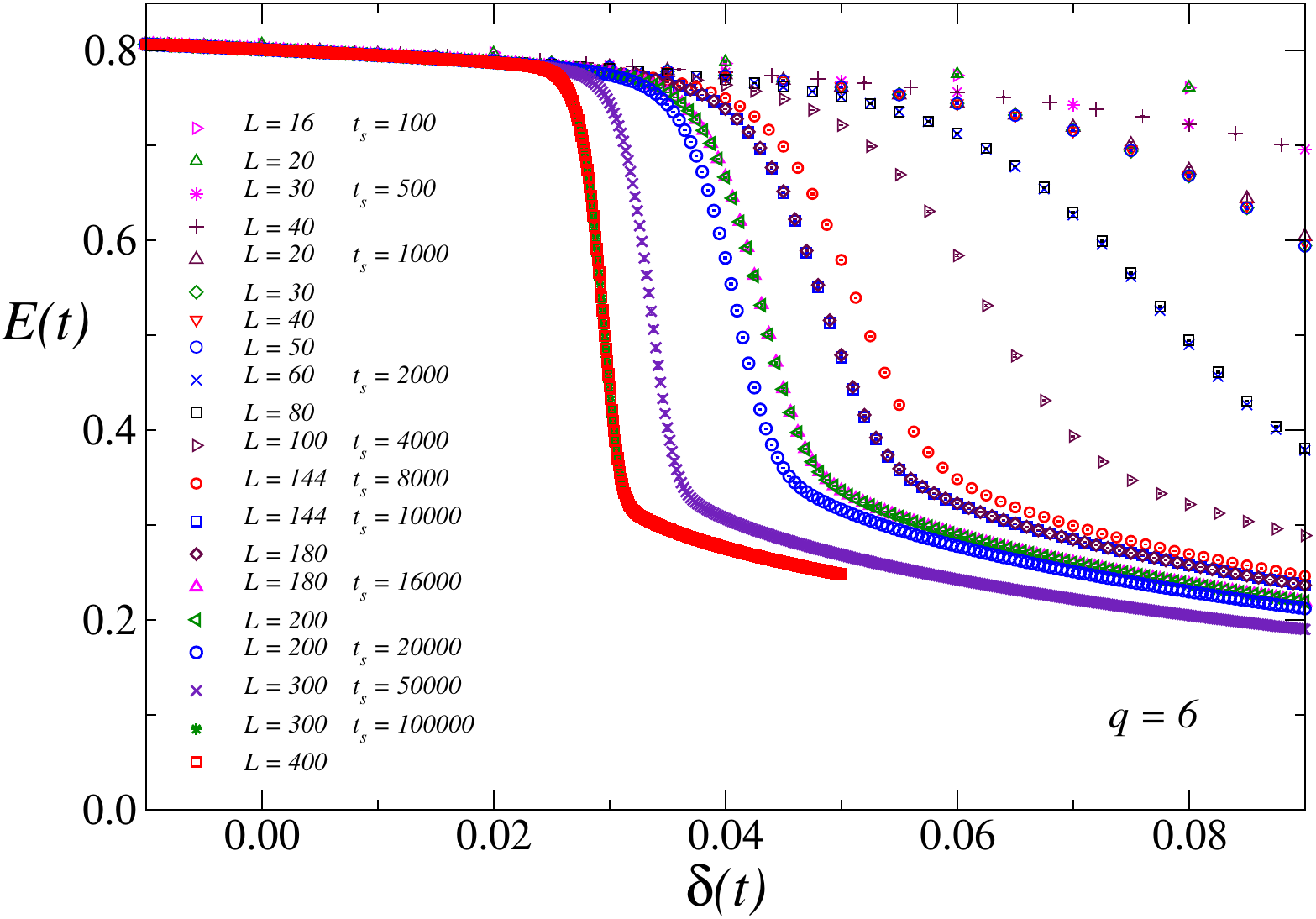}
  \caption{The time evolution of the energy density $E(t,t_s,L)$ for
    $q=6$, versus $\delta(t)=t/t_s$ for various values of $t_s$ and
    sizes $L$.  The comparison of data for different sizes and same
    $t_s$ (when $t_s$ is not explicitly reported, it is understood
    that its value is the one comparing in the symbol lines above)
    shows that the thermodynamic limit is effectively obtained for
    $L\gtrsim \sqrt{t_s}$ within the very small statistical errors of
    our simulations.}
\label{rawltsq6}
\end{figure}

\begin{figure}[tbp]
  \includegraphics[width=0.9\columnwidth, clip]{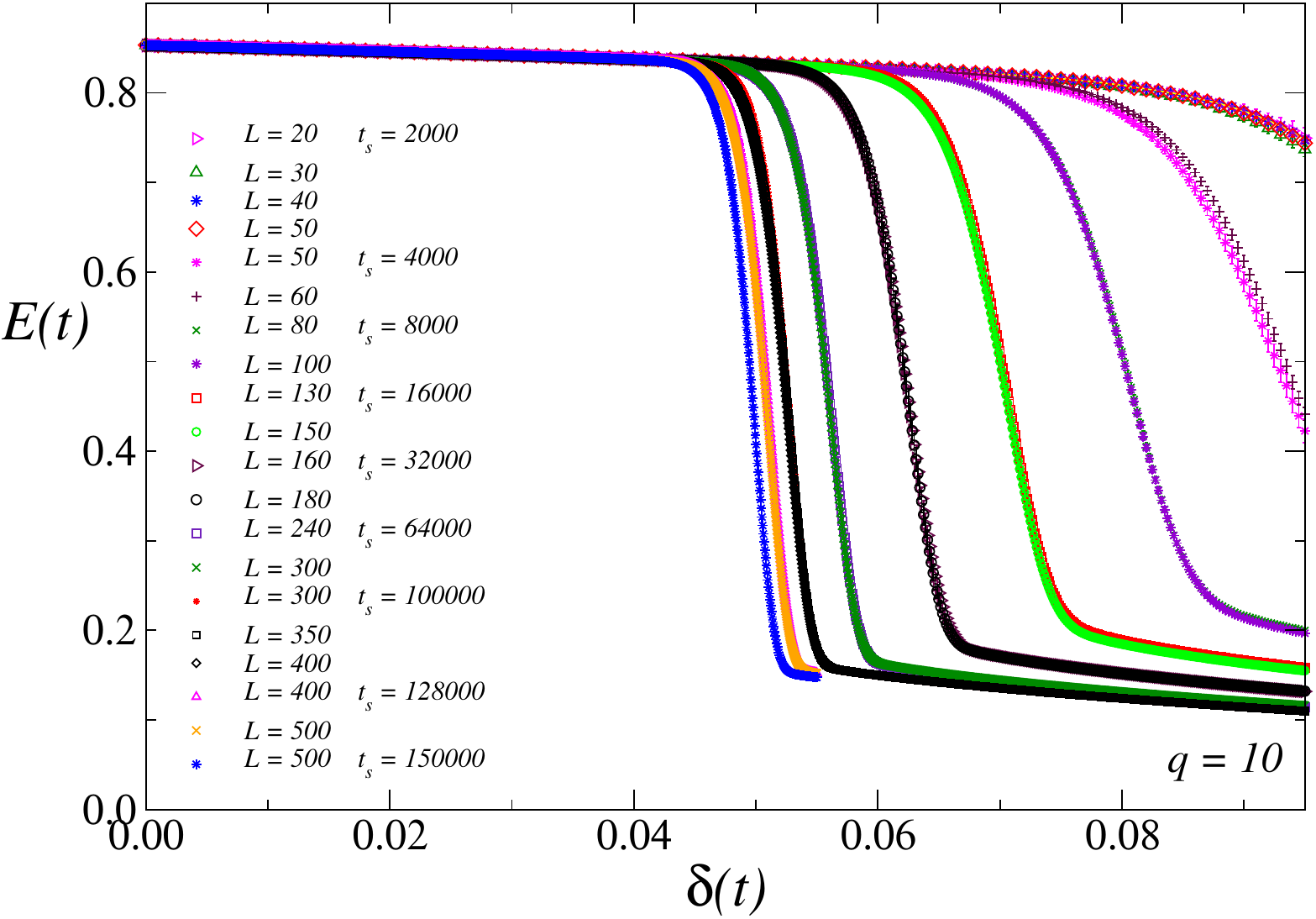}
  \caption{The time evolution of the energy density $E(t,t_s,L)$ for
    $q=10$, versus $\delta(t)=t/t_s$ for several values of $t_s$ and
    $L$.  Comparing data at different $L$ for the same $t_s$ indicates
    that the thermodynamic limit is reached for $L\gtrsim
    \sqrt{t_s}$.}
\label{rawltsq10}
\end{figure}

Figures~\ref{rawltsq6} and \ref{rawltsq10} show the time evolution of
the energy density versus $\delta(t)$, defined in Eq.~(\ref{deltat}),
for $q=6$ and $q=10$, respectively.  The comparison of the data for
different sizes and same $t_s$ shows that the thermodynamic limit is
effectively obtained for $L\gtrsim \sqrt{t_s}$ within the very small
statistical errors of our simulations (the relative errors of the data
when $E=0.5$ are typically less than one per cent for $q=6$, and at
most a few per cent for $q=10$). These results provide a good
approximation of the infinite-size limiting curve, within the accuracy
of the data.  Therefore, MC simulations of systems with lattice sizes
$L\lesssim 500$ allow us to collect data for the KZ dynamics in the
thermodynamic limit up to time scales $t_s$ of the order of $10^5$.
We mention that the energy density along the KZ trajectories shows
self-averaging with increasing lattice size, so a small number of
trajectories (of the order of few tens) is sufficient to provide
accurate results for the largest considered lattices.

\begin{figure}[htbp]
\includegraphics[width=0.9\columnwidth, clip]{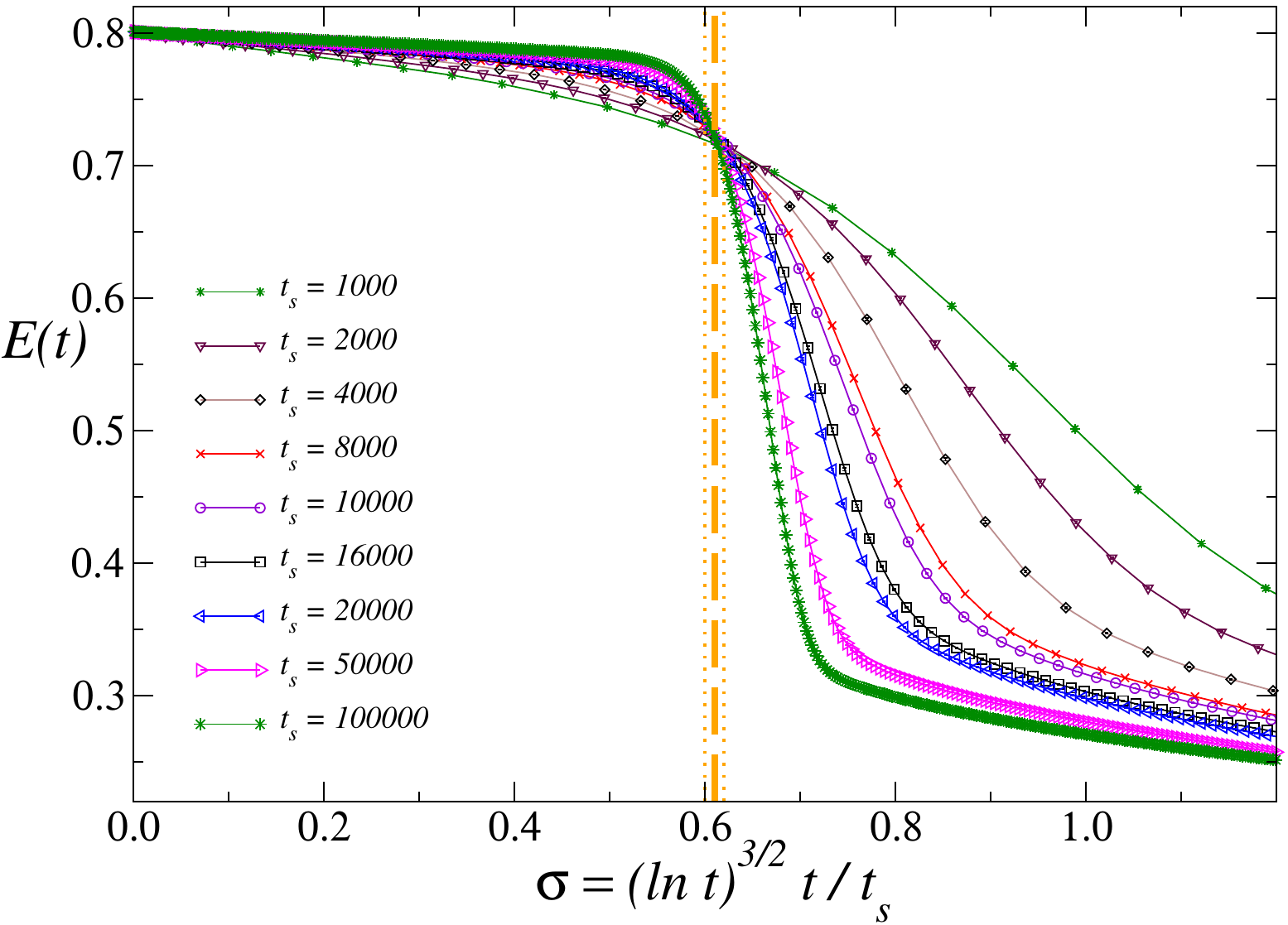}
\caption{Data for the $q=6$ energy density in the thermodynamic limit
  versus $\sigma=t (\ln t)^{3/2} /t_s$.  The vertical dashed line indicates
  the estimate $\sigma_*$ of the asymptotic crossing point; the interval 
  between the dotted lines gives the uncertainty.}
\label{firstrescl3o2}
\end{figure}

\begin{figure}[htbp]
    \includegraphics[width=0.9\columnwidth, clip]{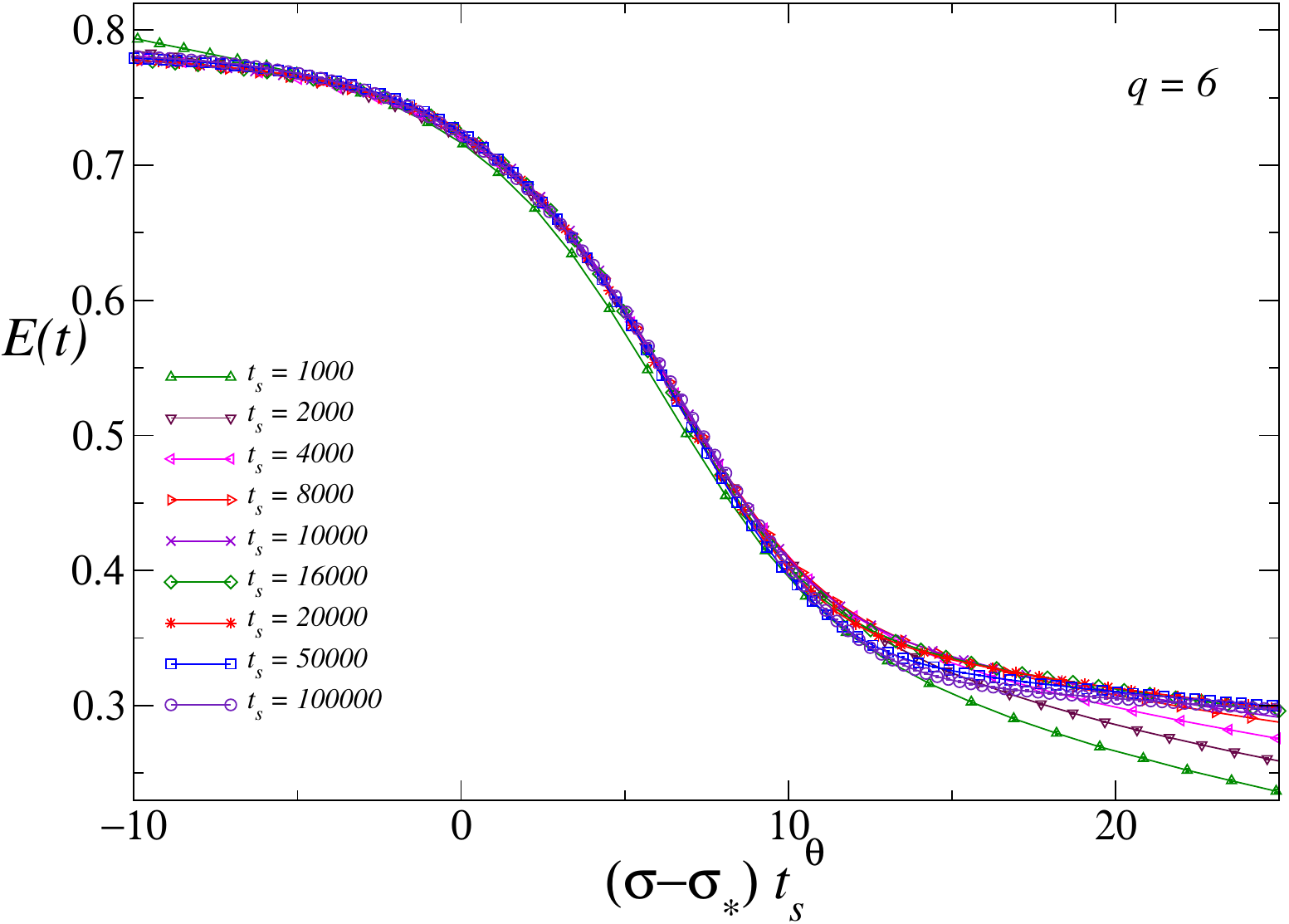}
\caption{The energy density $E(t)$ versus $\hat\sigma \equiv (\sigma -
  \sigma_*)\,t_s^\theta$ for $q=6$, using the optimal values
  $\sigma_*=0.61$ and $\theta=0.42$. The data appear to approach an
  asymptotic scaling curve with increasing $t_s$.}
\label{secondresc}
\end{figure}

Once the infinite-volume scaling curves for the energy are obtained,
we look for a scaling behavior in the large $t_s$ limit.
We first consider the $q=6$ Potts model. The data appear to scale in terms
of the variable $\sigma$ defined in Eq.~\eqref{sdef} with $\kappa=3/2$,
indicating that droplet nucleation is the relevant phenomenon in the
large-$t_s$ limit, as outlined in Sec.~\ref{TLsca}. Moreover, as shown in 
Fig.~\ref{firstrescl3o2}, the energy scaling function is singular 
(discontinuous) for $\sigma = \sigma_* \approx 0.61$. Indeed, all curves
have a stable crossing point at $\sigma_*$ and become increasingly steep
near this point as $t_s$ increases. This behavior is consistent with
the arguments of Sec.~\ref{TLsca} and reflects the fact the transition
between the two phases is much faster than the typical time required
to nucleate stable droplets.

The presence of a stable crossing point suggests the
emergence of an additional scaling behavior of the energy density versus
\begin{equation}
  \hat{\sigma}=(\sigma-\sigma_*) t_s^{\theta},
  \label{barsigma}
\end{equation}
for $\sigma$ close to $\sigma_*$.  This is confirmed by the data 
shown in Fig.~\ref{secondresc}, where we used
the optimal values $\sigma_*=0.61$ and $\theta=0.42$ (the relative  
uncertainty on $\sigma^*$ and $\theta$ is of a few percent and approximately
10\%, respectively).  The scaling in terms of $\hat{\sigma}$ 
shows that the asymptotic spinodal-like discontinuity is 
approached with corrections that decay as $t_s^{-\theta}$ as $t_s$ increases.
In the present
case we have $\theta = 0.42$, so corrections decay slowly.

\begin{figure}[htbp]
\includegraphics[width=0.9\columnwidth, clip]{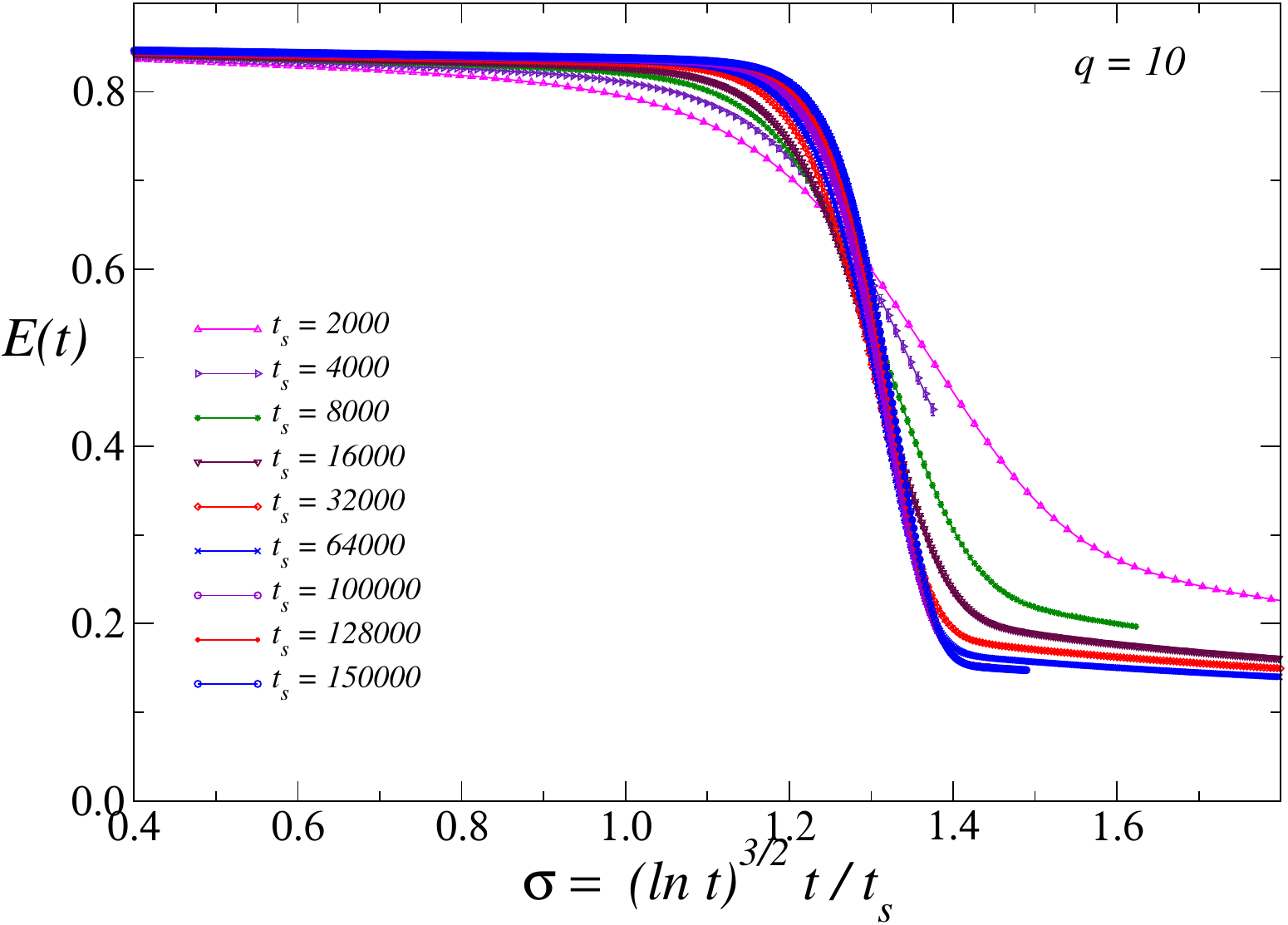}
\caption{Data for the energy density along the KZ protocol in the
  thermodynamic limit for $q=10$, versus $\sigma=t (\ln t)^{3/2}
  /t_s$. They show a good evidence of scaling with increasing $t_s$,
  and a sharp transition from the disordered phase to the low-temperature
  phase.}
\label{firstrescl3o2q10}
\end{figure}

We have also considered the Potts model for $q=10$. In this case as well, 
there is strong evidence that the energy data satisfy the 
scaling equation~\eqref{etsca} when taking $\kappa = 3/2$ (see 
Fig.~\ref{firstrescl3o2q10}). Furthermore, the transition from the
disordered to the ordered phase is very sharp, suggesting again
a discontinuity in the large-$t_s$ limit. Thus, for $q=10$ we again observe
a spinodal-like behavior, consistently with the idea that the relevant
time scale is associated with droplet nucleation. The transition between
the two phases is instead much faster.
The data for different time scales $t_s$ are again consistent with a
crossing point at $\sigma\approx 1.3$. However, the energy changes so
rapidly as a function of $\sigma$, that it is difficult to obtain
sufficiently precise estimates of $\sigma_*$ and $\theta$. Thus, we are not
able to reliably test the scaling in terms of the variable $\hat{\sigma}$
defined in Eq.~\eqref{barsigma}. 

We finally mention that we have also determined the range of $\kappa$
values consistent with the scaling behavior~\eqref{etsca}, to assess
the robustness of our scenario in which droplet nucleation governs the
observed spinodal-like scaling. From our analysis, we estimate an
uncertainty of approximately 10\% on the value $\kappa=3/2$.

\section{Conclusions}
\label{conclu}

We have reported a numerical study of the out-of-equilibrium
spinodal-like scaling behavior occurring in 3D $q$-state Potts models in
the thermodynamic limit, when the system is driven across the 
thermal FOT via a KZ relaxational dynamics.
This study provides additional insight into these dynamical phenomena,
which have been previously investigated at the thermal FOTs of
2D Potts models~\cite{PV-17} and at the magnetic FOTs of classical
and quantum Ising models~\cite{PV-25,PRV-25,PRV-25b}. In particular, our 
simulations are motivated by the somewhat puzzling results reported in
Ref.~\cite{PV-25} for the 3D and 4D Ising model. Specifically, 
while all 2D results, both for the Ising and Potts models,
can be explained by the simple assumption that droplet nucleation provides
the longest time scale for the transition between the two phases,
the 3D and 4D Ising results are significantly different, indicating
the presence of a distinct, longer time scale controlling the passage
between the two phases. This naturally raises the question of whether
the nucleation argument is valid only in two dimensions---hence it should not
work for the 3D Potts model---or whether the discrepancy arises from
some specific property of the Ising magnetic FOTs.

Our results for the 3D $q$-state Potts models with $q=6$ and $q=10$
confirm that the time-dependent energy density exhibits the scaling
behavior $E(t)\approx {\cal E}(\sigma)$, with $\sigma= { t \,(\ln
  t)^\kappa/t_s}$, as the scaling variable defined in
Eq.~\eqref{sdef}, and $\kappa=3/2$.  The estimated value $\kappa=3/2$
agrees with the value obtained by assuming that droplet nucleation is
the relevant mechanism in the large-$t_s$ limit, as explained in
Sec.~\ref{TLsca}.  Therefore, the numerical results confirm the
hypothesis that droplet nucleation sets the longest time scale for the
passage from the disordered to the ordered phase, whereas the
subsequent growth of such droplets occurs on a shorter time
scale. This framework also predicts that the scaling function ${\cal
  E}(\sigma)$ is discontinuous at a finite value $\sigma=\sigma_*$, a
feature that is further corroborated by our numerical findings.

Note that the value $\sigma_*$ where ${\cal E}(\sigma)$ is singular
corresponds to a $t_s$-dependent deviation $\delta\beta_*>0$ from the
transition point, which vanishes as
\begin{equation}
  \delta\beta_* \equiv \beta_* - \beta_{\rm fo}
    \sim {1\over (\ln t_s)^{3/2}}
\label{debe*}
\end{equation}
in the large-$t_s$ limit.  The observed scaling behavior resembles the
spinodal behavior predicted at FOTs by mean-field
calculations~\cite{Binder-87}.  However, there is a crucial
difference: in the mean-field case, the singular spinodal point
$\delta\beta_{\rm sp}>0$ remains different from zero in the
infinite-volume limit, while for the KZ dynamics $\delta\beta_*$
vanishes logarithmically for $t_s\to \infty$.  For this reason, we
refer to the out-of-equilibrium scaling behavior observed at the FOTs
of the 3D Potts models as spinodal-like.

Our results for the Potts model provide an example of a 3D system in
which droplet nucleation is the relevant mechanism controlling the
time behavior of the system for large values of $t_s$.  In contrast,
the scaling behavior observed at the magnetic low-temperature FOTs in
3D and 4D Ising systems~\cite{PV-25} indicate $\kappa=1$ and
$\kappa=1/2$ respectively, which differ from the values $\kappa = 3/2$
and $\kappa= 4/3$ predicted by the arguments of Sec.~\ref{TLsca}.
Clearly, another, as yet unidentified mechanism provides the largest
time scale for the passage from one magnetic phase to the other. For
instance, in Ising systems, droplets might have a fractal boundary, or
the relevant scale could be that associated with the transition time
from one phase to the other one, once stable droplets are created.
Further investigations are therefore necessary to achieve a complete
understanding of the out-of-equilibrium dynamics across FOTs,
particularly for magnetic FOTs occurring in 3D systems.

\end{document}